\documentclass[pageno]{jpaper}

\newcommand{\ignore}[1]{}
\usepackage[normalem]{ulem}
\usepackage{listings}
\usepackage{amsmath}
\usepackage[noend,linesnumbered,ruled,vlined]{algorithm2e}
\usepackage{physics}
\usepackage{amsmath}
\usepackage{multirow}
\usepackage{enumitem}
\usepackage{amssymb}
\usepackage{adjustbox}
\usepackage[normalem]{ulem}
\usepackage[flushleft]{threeparttable}
\usepackage{dblfloatfix}

\usepackage{graphicx}
\usepackage{subfig}

\usepackage{authblk}

\SetCommentSty{mycommfont}

\def\BibTeX{{\rm B\kern-.05em{\sc i\kern-.025em b}\kern-.08em
    T\kern-.1667em\lower.7ex\hbox{E}\kern-.125emX}}

\usepackage{url}

\usepackage[noadjust]{cite}

\begin{document}
\vspace{-1in}
\title{Are LLMs Good For Quantum Software, Architecture, and System Design?} 
\author[]{Sourish Wawdhane}
\author[]{Poulami Das}
\affil[]{The University of Texas at Austin}
\affil[]{\textit {\{sourishw, poulami.das\}@utexas.edu}}
\date{}

\maketitle

\thispagestyle{empty}

\section{Introduction}

Quantum computers promise massive computational speedup for problems in many critical domains, such as physics, chemistry, cryptanalysis, healthcare, etc~\cite{ibmqproteinfolding,lloyd1996universal,shor1994algorithms}. However, despite decades of research, they remain far from entering an era of \textit{utility}. The lack of mature software, architecture, and systems solutions capable of translating quantum-mechanical properties of algorithms into physical state transformations on qubit devices remains a key factor underlying the slow pace of technological progress. Algorithmic research at the top, such as designing efficient algorithms to map applications or building quantum error correction (QEC) protocols to tolerate device errors, are largely limited to theoretical studies. In contrast, device level research at the bottom is either driven by industry that focus on a single qubit modality or academic labs that over-optimize qubit devices at a small scale; however, these optimizations are often fragile and tend to break down as system sizes increase. The problem worsens due to significant reliance on domain-specific expertise, especially for software developers, computer architects, and systems engineers. To address these limitations and accelerate large-scale high-performance quantum system design, we ask: \vspace{0.1in} \\ \textit{\textbf{Can large language models (LLMs) help with solving quantum software, architecture, and systems problems?}}

\section{Background: A Case Study}

To answer this question, we perform a preliminary case study. We are the instructors (primary instructor and teaching assistant) for the "Introduction to Quantum Computing Systems" course at UT Austin~\cite{ece382v}. This course: 
(1)~introduces quantum computing basics to ECE/CS students, (2)~familiarizes them with on-going research in the field (with particular emphasis on programming, compiler optimizations, architectural and systems models, based on recently published research papers at premier IEEE and ACM conferences, such as ISCA, MICRO, ASPLOS, HPCA, QCE), and (3)~trains them to optimize quantum software and architecture. A critical component of the course focuses on in-class exams in which the students use their existing knowledge to solve quantum system design problems. The exam duration is one hour and thirty minutes and students are allowed to bring in calculators and a cheat-sheet written by themselves (although the students generally claim that cheat sheets are not so helpful for them to attempt the questions). For our case study, we take one of these exams and study the performance of different LLMs from OpenAI, Google, and Anthropic. Table~\ref{tab:benchmarks} offers a detailed summary of these model families and versions.

\begin{table}[htp]
\renewcommand{\arraystretch}{1}
\setlength{\tabcolsep}{5pt}
\caption{Summary of LLMs used. For each provider and model family, we use two variants- a lightweight version for producing fast outputs and an advanced version with reasoning abilities.}
\centering
\begin{tabular}{|c|c|c|c|}
    \hline

Provider & Model Family & Lightweight & Reasoning\\
\hline
    \hline
OpenAI & GPT \cite{openai2024models} & 5.3 & 5.4-Thinking\\
\hline
Google & Gemini \cite{team2024gemini} & 3 Fast & 3.1-Pro \\
\hline
Anthropic & Claude \cite{anthropic2024claude} & Sonnet 4.6 & Opus 4.6 \\
\hline

\end{tabular}
\label{tab:benchmarks}
\end{table}

\ignore{
\begin{table}[htp]
\renewcommand{\arraystretch}{1.2}
\setlength{\tabcolsep}{2pt}
\caption{Summary of LLMs used in our case study}
\small
\begin{tabular}{|l|c|c|c|}
    \hline
Provider & Model Family & Version & Key Features \\
\hline
\multirow{3}{*}{OpenAI} & \multirow{3}{*}{GPT} & 5.3 & Speed, cognitive density \\
\cline{3-4}
& & 5.4-Thinking & Deeper reasoning,\\
& & & accuracy, precision\\
\cline{1-4}
\multirow{2}{*}{Google} & \multirow{2}{*}{Gemini} & 3 Fast & Speed\\
\cline{3-4}
& & 3.1-Pro & Advanced reasoning\\
\cline{1-4}
\multirow{4}{*}{Anthropic} & \multirow{4}{*}{Claude} & Sonnet 4.6 & High-performance, \\
& & & 
cost-effective\\
\cline{3-4}
& & Opus 4.6  & Better thinking,\\
& & & comprehensive responses \\
\hline

\end{tabular}
\label{tab:benchmarks}
\end{table}}

The exam covered topics related fault-tolerant quantum systems, including trade-offs in QEC codes, decoder designs~\cite{alavisamani2024promatch,das2022afs}, synchronizing QEC cycles~\cite{maurya2025synchronization}, and designing flag-proxy networks~\cite{vittal2024flag}. To compare the performance against human \textit{experts}, we show the test scores of four student participants. Note that these students willingly participated in the study and FERPA laws are not violated. 
The performance of the models and students were both manually graded by the instructors. To measure performance, we use the exam score out of a total of 100 points. The exam, rubric, solutions, and exam scores are available at \url{https://tinyurl.com/arellmsgoodatquantum}.

\begin{figure*}[t]
    \centering
    \includegraphics[width=\linewidth]{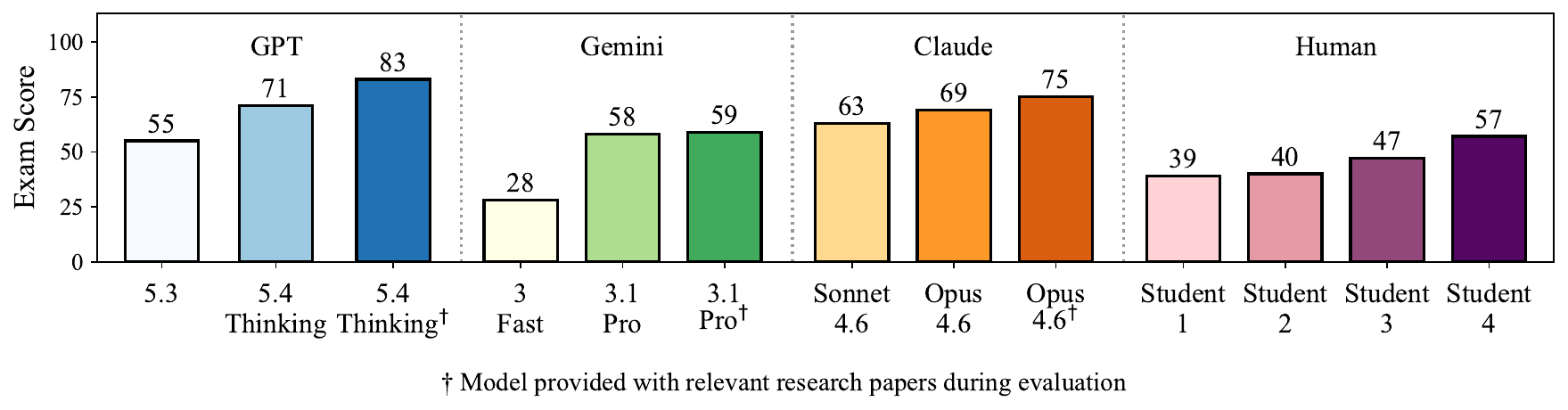}
    \caption{Performance of different large language model families for quantum reasoning tasks, especially related to software, architecture, and system design. Overall, all the model families perform well on the exam, especially the reasoning variants of each model. The performance also increases when assisted by a human expert.}
    \label{fig:results}
\end{figure*}

\section{Preliminary Results}

Figure~\ref{fig:results} shows the performance (exam score) for different model configurations and compares against the performance of the students. We make the following observations.

\begin{enumerate}[leftmargin=0cm,itemindent=.5cm,labelwidth=\itemindent,labelsep=0cm,align=left, itemsep=0.1 cm, listparindent=0cm]
    \item All six LLMs used performed well with an average score of $57.33$. The minimum is $28$ using the Gemini-3 Fast model while the maximum is $71$ using the GPT-5.4 Thinking model.  
    \item Reasoning models outperform lightweight variants with an average score of $66$ compared to $48.67$. Reasoning models increase scores by $1.3\times$ and $2.1\times$ respectively for the GPT and Gemini models, compared to only $1.1\times$ for the Claude models. GPT 5.4 Thinking used the largest reasoning traces (1m 57s without papers and 4m 18s with papers). Gemini Pro used 56s without papers and 1m 23 seconds with papers. In contrast, Claude reasoning models chose not to reason.
    \item Assisting the reasoning LLMs to use relevant research papers~\cite{alavisamani2024promatch,maurya2025synchronization,vittal2024flag} increases the average score further to $72.33$. The improvements are much more pronounced for the GPT and Claude models than the Gemini model. 
    \item The GPT models outperform the other two LLM families for each configuration and attains the overall highest score of $83$, outperforming the students by a significant margin.
    \item LLMs performed poorly on a question requiring test-takers to map QEC codes to hardware. LLMs struggled to discover optimal mappings and reason about error masking. 
\end{enumerate}

\section{Future: Quantum Systems Aided by AI}

Based on this preliminary case-study, we think research at the quantum and AI frontier is a promising and critical direction.

\vspace{0.1in}
\noindent \textbf{A Promising Outlook:} The overall performance of LLMs on the exam (which the instructors as well as the students consider to be significantly difficult) is promising and highlights the possibility of using LLMs to accelerate quantum software, architecture, and systems development. A fundamental research question on this front remains: \textit{how do we build agentic workflows specifically for quantum system design?}

\vspace{0.10in}
\noindent \textbf{Creating Quantum Specific LLMs:} Modern day LLMs are primarily designed for coding tasks, content creation, summarization, Q\&A settings. However, we lack LLMs specifically trained for quantum reasoning tasks. Creating dedicated models or fine-tuning existing LLMs have the potential to improve performance even further. We should spend research and engineering efforts into this stream to effectively use LLMs for quantum reasoning tasks. In particular, all the models used in our study show excellent performance in error decoder design questions in the context of QEC and surface codes. Accurate, fast, and scalable decoders are critical to enable QEC at scale. Using quantum-specific LLMs have the potential to further assist in their development.

\vspace{0.10in}
\noindent \textbf{Need for Training Datasets:} We understand that the current case-study is done in a very limited setting and requires more involved expert participation to refine the scope and assess the headroom for improvement. It also requires advanced training recipes to steer the model to generate more accurate and nuanced outputs. For example, a similar study designed and led by researchers from Harvard University along with participants from various top-ranked computer engineering programs \cite{prakash2025quarchquestionansweringdatasetai} show that using curated datasets for computer architecture problems improve the reasoning capability of traditional LLMs. We anticipate a similar methodology to remain effective even for quantum reasoning tasks. We recommend concerted efforts in this space to expand the capabilities of LLMs for quantum tasks.

\vspace{0.10in}
\noindent \textbf{Role of Human Experts Remain Critical:}
The performance of all three model families used in our case study improves when the reasoning models are assisted by the instructor to use the right research paper. Especially in a field like quantum computing and topics involving software, architecture, and systems design, the human expert still plays a critical role in the adoption of LLMs. It also remains unclear if moving to larger models with more parameters and advanced reasoning capabilities can close the gap by reducing the reliance on the expert or we would hit a parameter wall. This also poses a more crucial question- how do we maximize the utility of the human experts given the limited workforce in this domain and the criticality of their inputs.

\vspace{0.10in}
\noindent \textbf{LLMs Struggle On Certain Advanced Topics:}
All the LLMs performed poorly on a quantum reasoning task designed around optimal design of flag-proxy networks. Flag proxy networks optimally use flag and proxy qubits to reduce error propagation and overcome the limited connectivity of superconducting quantum systems. For higher resource efficiency and error performance, designing these networks with minimal qubits is critical. Six models failed to generate valid flag connectivity configurations that minimized flag usage or required swaps. Moreover, all the LLMs under study were unable to reason about multi-error scenarios, where error propagation can mask syndromes via interference.  These findings suggest that current LLMs struggle with certain complex reasoning tasks, motivating the use and improvement of multi-modal LLMs to solve these problems with higher accuracy.

\section{Conclusion}
LLM-assisted workflows and AI-based frameworks promise revolutionary impact in traditional software ecosystems. In this paper, we show how LLMs can also be used to power quantum software, architecture, and systems development by evaluating the performance of traditional LLMs on quantum reasoning tasks. Based on our study, we have recommended several directions along which research and engineering development efforts must be pursued.

\section*{Acknowledgments}
We thank Shashank Nag, Allison Seigler, Dongwhee Kim, and Pawan Kashyap, for their participation in this study. We thank Avinash Kumar for generating Claude responses. We have open sourced the exam as well as the rubric and we encourage others to contribute to the dataset as well.

\bibliographystyle{unsrt}
\bibliography{references}

\end{document}